\documentclass[epj-spec]{svjour}

\usepackage{graphics,amsmath,amssymb,epsfig,color}
\begin{document}
\title{Transport in quasi one-dimensional  spin-1/2 systems}
\author{F. Heidrich-Meisner\inst{1}\fnmsep\thanks{\email{fheidric@utk.edu}} \and A. Honecker\inst{2} \and W. Brenig\inst{3} }
\institute{Materials Sciences and Technology Division, Oak Ridge National Laboratory,  
Tennessee 37831, and Department of Physics,  University of Tennessee, Knoxville, Tennessee 37996, USA\and Institut f\"ur Theoretische Physik, Universit\"at G\"ottingen,
37077 G\"ottingen, Germany \and Technische Universit\"at Braunschweig, Institut f\"ur Theoretische
  Physik, Mendelssohnstrasse 3, 38106 Braunschweig, Germany}
\abstract{
We present numerical results for the spin and thermal conductivity of
one-dimensional (1D) quantum spin systems. We contrast the properties of 
integrable models such as the spin-1/2 $XXZ$ chain against nonintegrable ones such as
frustrated and dimerized chains. The thermal conductivity of the $XXZ$ chain 
is ballistic at finite temperatures, while in the nonintegrable models, this quantity is argued
to vanish. For the case of frustrated and dimerized chains, we discuss 
the 
 frequency dependence of the transport coefficients. Finally, we give an overview over related theoretical 
 work on intrinsic and extrinsic scattering mechanisms of quasi-1D spin systems.
} 
\maketitle
\section{Introduction}
\label{intro}

Quantum magnetism in  1D is a successful example for a fruitful interplay between
theory and experiment. On the one hand, many   bulk materials exist that almost perfectly 
realize 1D spin models
(see \cite{dagotto99,johnston00} for a review) and on the other hand, powerful theoretical methods
such as bosonization \cite{schulz96}, the Bethe ansatz \cite{bethe31}, or the density-matrix renormalization group method \cite{white92b}
are available. Often, excellent agreement between theoretical predictions 
and experiments has been found
as far as    ground-state properties, excitation spectra, thermodynamic or optical properties are concerned (see
\cite{dagotto99,qm-review}).
The understanding of transport properties   is of great importance for the interpretation of 
transport or NMR measurements (see, e.g., Refs.~\cite{takigawa96}). Substantial progress has been made
in the past years, but the understanding 
is still incomplete, especially for systems involving many coupled degrees of freedom such as 
spins, orbitals, and phonons.    
Theoretically, the topic of transport in 1D quantum magnets is challenging. First, several experiments demand for a more complete theoretical picture as 
will be outlined below, and second, transport theory often requires the computation
of non-trivial correlation functions \cite{mahan}.
Third, transport is closely related to relaxation and non-equilibrium phenomena and thus connects
to the rapidly evolving field of  non-equilibrium physics of strongly correlated electron systems. 

In particular, the discovery of the {\it colossal} 
magnetic heat transport
in spin ladder materials such as (Sr,Ca,La)$_{14}$Cu$_{24}$O$_{41}$, where the magnetic contribution to the 
total thermal conductivity $\kappa$ exceeds the phonon part substantially \cite{solo00,hess01,kudo01}, has sparked
 interest in transport properties of quasi-1D spin models. Often,  
 a magnetic mean-free path is defined within a Boltzmann type of description \cite{solo00,hess01} used to analyze the experimental data, 
which in the case of spin ladders can be of the order of several hundred lattice constants \cite{hess01}. This observation -- originally
suggested to reflect  ballistic transport properties of pure spin ladders \cite{alvarez02b} -- 
is  not yet completely understood. Several other spin-1/2 chain materials (see, e.g., \cite{solo01,hess07}) and 
2D cuprate antiferromagnets \cite{sales02,sun03}
possess similar thermal transport properties, although typically the values measured for the magnetic contribution are much
smaller. A very interesting aspect is the strong magnetic field dependence observed in some 2D \cite{sales02,hofmann01} and 1D materials
\cite{ando98,sologubenko07}.
Not all materials that exhibit a strong dependence of $\kappa$ on magnetic
field are actually believed to have a significant contribution to $\kappa$
from magnetic excitations. Nevertheless, the possibility of tuning
the thermal current through magnetic fields is appealing and may even
allow to design functional devices such as spin valves \cite{sales02}.   
 We refer the reader to a recent review \cite{sologubenko07a} and the article by C. Hess in this volume 
for more details on the experimental developments.

Much theoretical work has  focused on intrinsic transport properties of spin systems, addressing intriguing  questions such as the different
transport properties of integrable as compared to nonintegrable ones.
 While in the remainder of this article, we restrict the discussion to the application
of linear response theory -- i.e., Kubo formulae -- to spin and thermal transport of quasi-1D spin-1/2 systems as 
derived in Refs.~\cite{mahan,luttinger64}, we  note that 
alternative approaches such as master-equation techniques incorporating a modeling of heat baths have been pursued for quantum 
systems \cite{saito96,mejia-monasterio06,michel03}. 
Moreover, while widely used, the derivation of Kubo formulae for heat transport may be questioned, as strictly
speaking, no analogue to the voltage or magnetization gradients driving electrical and spin currents exists
in the case of thermal transport. We refer the reader to recent work on this issue \cite{michel03,michel04,michel05a,gemmer06}.
For brevity, we also concentrate on spin-1/2 systems and refer to the literature for more details on Haldane systems
\cite{sachdev97}. Note, though, that due to similar low-energy properties \cite{hida92}, the transport
behavior  of gapped quantum systems such as spin ladders and spin-1 chains can be expected to be  generic
at low temperatures.
Analogous questions, i.e., the properties of integrable vs
nonintegrable systems, the validity of Fourier's law, and the modeling of heat baths are timely subjects in the study
of transport of classical systems (see Ref.~\cite{lepri03} for a review).

In linear response theory, ballistic transport is defined by the existence of a finite Drude weight
$D$ \cite{kohn64,scalapino93},
which is the zero-frequency contribution to the real part of the conductivity:
\begin{equation}
\mbox{Re}\, \kappa[\sigma](\omega) = D_{\mathrm{th[s]}} \delta{(\omega)}  +\kappa[\sigma]_{\mathrm{reg}}(\omega) \label{eq:defineD}\,,
\end{equation}
where $\kappa$ denotes the thermal and $\sigma$ the spin conductivity. $\delta(\omega)$ is a $\delta$-function and
$\kappa[\sigma]_{\mathrm{reg}}(\omega)$ is assumed to be regular at $\omega=0$.
Generally, the transport coefficients are computed from current-current correlation functions:
\begin{equation}
\kappa[\sigma](\omega)  
= -\frac{\beta^{r}}{N} \int_0^{\infty} dt\, e^{i(\omega+i0^+ )t}
         \int_0^{\beta} d\tau \langle j_{\mathrm{th[s]}}\,j_{\mathrm{th[s]}}(t+i\tau)\rangle \, .
\end{equation} 
 Here and in all succeeding equations, $r=0$ for spin transport (labeled by 's') and $r=1$ for thermal transport (labeled by 'th'). $\beta=1/T$ is the 
inverse temperature and $\langle .\rangle$ denotes the thermodynamic expectation value. 
A finite Drude weight implies a divergent dc conductivity. If $D$ vanishes, then either a finite dc 
conductivity
$\sigma_{\mathrm{dc}}=\lim_{\omega\to 0}\sigma_{\mathrm{reg}}(\omega) $
can result, or, if  $\sigma_{\mathrm{reg}}(\omega)$ exhibits 
an anomalous frequency dependence for $\omega\to 0$, $\sigma_{\mathrm{dc}}$ may still diverge \cite{zotos-review}. Note that here, we mainly
consider finite temperatures, while the Drude weight was original introduced by Kohn to characterize a  metal at $T=0$ 
\cite{kohn64,scalapino93}.

Trivially, if the respective current operator commutes
with the Hamiltonian, the Drude weight is finite at
any temperature. It has long been known that the energy current operator of the spin-1/2 $XXZ$ chain is a conserved quantity
\cite{niemeyer71},
but only  later, a deeper connection between the existence of finite Drude weights at finite temperatures and the {\it integrability}
of a model system has been made \cite{zotos97}.

As a main objective of this paper,   we wish to 
summarize recent theoretical progress, concentrating on one-dimensional systems and their 
intrinsic spin and heat transport properties (see  also Refs.~\cite{zotos-review,zotos05} for recent reviews). 
We will contrast the properties of integrable systems such as the spin-1/2 $XXZ$ chain discussed in Sec.~\ref{sec:2}
against nonintegrable ones. As an example for the latter class of systems, we present
numerical results for the spin and thermal conductivity of the  frustrated and
dimerized spin-1/2 chain in Sec.~\ref{sec:3}.  
With respect 
to the experimental findings, obviously, both {\it intrinsic} as well as {\it extrinsic} scattering 
processes are of relevance. Recent theoretical results on extrinsic scattering channels are summarized in Sec.~\ref{sec:ex}.

%

\section{Transport properties of the $XXZ$ chain}
\label{sec:2}
We now turn   to the nearest-neighbor spin-1/2 $XXZ$ chain. The Hamiltonian is:
\begin{equation}
H_{XXZ} =\sum_{l}h_l=J \sum_{l} \left\lbrack \frac{1}{2}(S_l^+ S_{l+1}^- +h.c. ) + \Delta S_l^z S_{l+1}^z\right\rbrack \,.
\label{eq:xxz}
\end{equation}
We set $J=1$ in the following and periodic boundary conditions are imposed throughout this work.
The current operators corresponding to the local energy density $d_l=h_l$  defined in Eq.~(\ref{eq:xxz}) 
and local spin density $d_l=S_l^z$ are obtained from the equations of continuity:
\begin{equation}
j_{\mathrm{th[s]},l+1}  - j_{\mathrm{th[s]},l}	=  -i\lbrack H, d_{\mathrm{th[s],l}}\rbrack \quad \Rightarrow \quad j_{\mathrm{th[s]}}=i\sum_{l=1}^N
\lbrack h_{l-1}, d_{l}\rbrack \,.
\label{eq:f2}
\end{equation}
It turns out that the energy current of the spin-1/2 $XXZ$ chain is a  nontrivial 
conserved quantity of this integrable model \cite{zotos97}.\footnote{{\it Trivial} conserved quantities are,
e.g., the total energy.} Hence in the case of thermal transport, 
$\mbox{Re}\, \kappa(\omega) = D_{\mathrm{th}} \delta{(\omega)}$ for any exchange anisotropy $\Delta$ of this model.
  Although the spin current is not in general conserved in the case of the spin-1/2 $XXZ$
chain, it has nevertheless been conjectured that $D_{\mathrm{s}}$ should be finite at $T>0$ in the case of integrable models, but vanish in the 
case of nonintegrable ones \cite{castella95,zotos96}. 
While we will  argue in Sec.~\ref{sec:3} that this picture seems to be correct for the massive phases of nonintegrable 1D spin models,
some counterexamples have been proposed in the literature \cite{kirchner99,heidarian07}. 

\subsection{The thermal and the spin Drude weight at zero magnetic field}
\label{sec:2.1}

\begin{figure}[t]
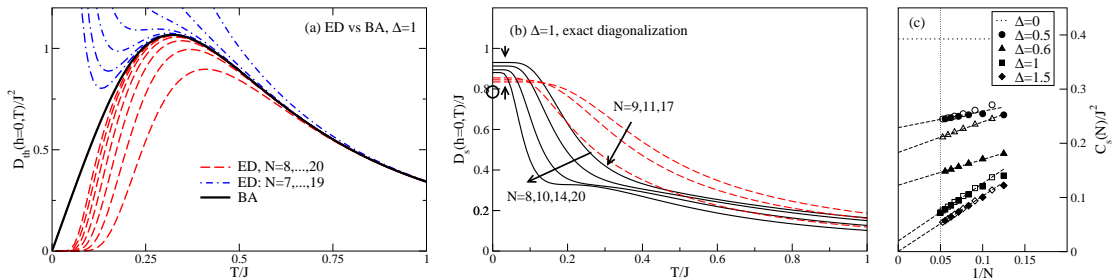

\centerline{ 
 \epsfig{figure=Dth_XX1_h0.eps,height=0.25\textwidth} \hspace{0.2cm}\epsfig{figure=Ds-xx1.eps,height=0.25\textwidth} \hspace{0.1cm}
\epsfig{figure=Csp_XXZ.eps,height=0.25\textwidth}}
\caption{Spin-1/2 $XXZ$ chain. (a):
Thermal Drude weight,  ED (dashed lines: even $N$, dot-dashed lines: odd $N$; see \cite{hm02}) vs 
BA results (solid line, \cite{kluemper02}) at $\Delta=1$.  (b): Spin Drude weight as a function of temperature and for 
several system sizes at 
$\Delta=1$. 
(c)  Finite-size scaling at high temperatures: $C_{\mathrm{s}}=\lim_{T\to\infty}\lbrack T \cdot D_{\mathrm{s}}\rbrack$ for 
$\Delta=0,0.5,0.6,1,1.5$ (from \cite{hm03}, with additional data for $N=20$ at $\Delta=1$). }
\label{fig:dsxxz}
\end{figure}

As for the Drude weights of the spin-1/2 $XXZ$ chain, the following picture has emerged: the thermal transport
is ballistic for any exchange anisotropy and at all
non-zero temperatures. Its dependence on $T$ and $\Delta$ has been 
studied by means of Bethe-ansatz (BA) techniques \cite{kluemper02,sakai03}, exact diagonalization (ED) \cite{alvarez02b,hm02,hm03},
and with mean-field theory \cite{hm02,lou04}.
An example is shown in Fig.~\ref{fig:dsxxz}(a), where we display ED data for the thermal Drude weight of the $XXZ$ chain at $\Delta=1$
vs temperature \cite{hm02}, in comparison with BA results from Ref.~\cite{kluemper02}. Note that numerically, the Drude weight can be computed
from \cite{zotos97} 
\begin{equation}
D_{\mathrm{th[s]}}(T) =
   \frac{\pi\beta^{r+1} }{Z\,N} \sum_{m, n \atop E_m=E_n} e^{-E_n/T}
    | \langle m | j_{\mathrm{th[s]}}|n\rangle |^2  . 
\label{eq:drude}
\end{equation}
Here, $ |n\rangle$ and $E_n$ are eigenstates and -energies of $H$, respectively, and $Z=\sum_n e^{-E_n/T}$
denotes the partition function.
Using system sizes as large as $N=20$, the ED agrees with  BA  down to temperatures of $T/J\sim 0.25$, which can be
improved by employing extrapolation methods \cite{hm02}.

Spin transport in the $XXZ$ chain is a more involved problem as
$[ H, j_{\mathrm{s}}]\not=0$, and has been the objective of many studies
\cite{zotos97,hm03,lou04,zotos96,narozhny98,fabricius98,naef98,zotos99,peres99,alvarez02,fujimoto03,long03,rabson04,laflorencie04,prelovsek04,benz05,gobert05,sirker06,mukerjee06,mukerjee07}.
While the spin Drude weight at $T=0$ is known exactly \cite{shastry90}, for the massless regime $|\Delta |<1$, where agreement
exists that $D_{\mathrm{s}}(T>0)>0$, several BA calculations arrive at contradicting results for the temperature
dependence \cite{zotos99,benz05,lan07}. The same holds for the question whether $D_{\mathrm{s}}(T>0)$ is finite or not at the 
SU(2) symmetric point $\Delta=1$. We refer the reader to Refs.~\cite{zotos05,benz05}
for a discussion of this issue
and the conceptual problems that BA approaches face. Recent numerical studies
are consistent with $D_{\mathrm{s}}(T>0)>0$ \cite{hm03,alvarez02,mukerjee07}.  We illustrate this in Fig.~\ref{fig:dsxxz}(b) and (c) where we show ED  results
for $D_{\mathrm{s}}(T>0)>0$ as a function of temperature in Fig.~\ref{fig:dsxxz}(b)  and a finite-size scaling analysis of 
the leading residue $C_{\mathrm{s}}=\lim_{T\to \infty}\lbrack T \cdot D_{\mathrm{s}}\rbrack$ in Fig.~\ref{fig:dsxxz}(c). Within
numerical precision and under the assumption that $C_{\mathrm{s}}\propto 1/N$ does not change at very large $N$, the
extrapolation results in finite values for $|\Delta |\leq 1$. We refer to Ref.~\cite{hm03} for a detailed discussion of
the finite-size scaling and to Refs.~\cite{hm03,peres99,prelovsek04} for recent work on the massive, antiferromagnetic regime
$\Delta>1$. Numerical results for the ferromagnetic phase $\Delta\leq -1$ can be found in Ref.~\cite{hm02}. The frequency dependence
of the regular part of the spin conductivity has numerically been studied  in Refs.~\cite{zotos96,naef98,prelovsek04,hm-diss}.

%

\subsection{Finite magnetic fields}
\label{sec:2c}
In the presence of magnetic fields $H_{Z}=-h \sum_i S_i^z$, the energy current $j_{\mathrm{th}}$ and the 
spin current  $j_{\mathrm{s}}$ couple since particle-hole symmetry is broken and hence $\langle j_{\mathrm{th}} j_{\mathrm{s}}\rangle
\not= 0$ \cite{zotos97}.\footnote{We invoke the notion of particle hole symmetry as 1D spin models
can be mapped onto spinless fermions with local interactions using the Jordan-Wigner
transformation \cite{mahan}.} This gives rise to magnetothermal effects, similar to the Seebeck effect of conduction electrons \cite{mahan}.
Transport in the $XXZ$ chain in the presence of magnetic fields has been studied theoretically in 
\cite{louis03,hm05,sakai05,furukawa05}, focusing on the magnetothermal effect and the thermomagnetic power.
To describe transport, now a  $2\times 2$ matrix description is necessary \cite{mahan,louis03,hm05}:
\begin{equation}\left(\begin{array}{c}
J_{1} \\
J_{2}
\end{array}\right) =
\left(\begin{array}{cc}
L_{11} & L_{12} \\
L_{21} & L_{22}
\end{array}\right) 
\left(\begin{array}{c}
\nabla h\\ 
-\nabla T
\end{array}\right)\,,
\label{eq:k0}
\end{equation}
where $\nabla h$ and $\nabla T$ are the gradient in field and temperature, respectively.\footnote{Note that $J_1=j_{\mathrm{s}}$ and
$J_2=j_{\mathrm{th}}-h j_{\mathrm{s}}$ and therefore, $D_{\mathrm{s}}=D_{11}$.}
The conservation of $j_{\mathrm{th}}$ is sufficient to show that all  four
 Drude weights $D_{ij}$ corresponding to the transport coefficients $L_{ij}$ are finite at all temperatures \cite{zotos97}.
 Then, assuming the condition of zero magnetization current flow, the thermal Drude weight $K_{\mathrm{th}}$ is obtained as
 \begin{equation}
 K_{\mathrm{th}}(h,T) = D_{22}(h,T) -  \frac{D_{12}^2(h,T)}{T\,D_{11}(h,T)}\,,
 \end{equation}
where now the magnetothermal correction $D_{12}^2(h,T)/\lbrack {T\,D_{11}(h,T)} \rbrack$ contributes as well.
Some of the main results of our work Ref.~\cite{hm05} are (i) a reduction of  $K_{\mathrm{th}}(h,T)$ due to the magnetothermal correction 
and (ii) expressions for the leading contributions to $K_{\mathrm{th}}(h,T)$ at low temperatures. The latter has been obtained
from mean-field theory and bosonization. In the gapless phase of the $XXZ$ chain (see, e.g., \cite{cabra98} for details on the phase
diagram), $K_{\mathrm{th}}(h,T)\propto T$, but at the quantum
critical line separating the gapless from the ferromagnetic regime, we find $K_{\mathrm{th}}(h,T)=A T\,^{3/2}$ with $A$ independent of
the exchange anisotropy $\Delta$. Note that $K_{\mathrm{th}}(h,T)$ has not been calculated yet by means of BA since 
$D_{\mathrm{s}}(h,T)$ escapes an analytical treatment \cite{benz05,sakai05}. Mean-field theory, as outlined in
Refs.~\cite{hm02,hm03,hm05}, proves useful as it provides a quantitatively good approximation to the thermal Drude weight both in 
zero \cite{hm02} and finite magnetic fields \cite{hm05}.

In most cuprate based spin chain and ladder materials, exchange couplings are of the order of 1000K \cite{dagotto99,johnston00} hence 
little effects of a magnetic field on the thermal conductivity have been observed. 
In a recent experiment, the thermal conductivity of  copper pyrazine dinitrate has been studied \cite{sologubenko07}. 
Since $J\sim 10.3K$ \cite{hammar99}, a significant
field dependence is found at low temperatures. The analysis of the experimental data employing the mean-field theory description of
Ref.~\cite{hm05} yields a constant mean-free path. The origin of this result especially at the quantum critical line remains to be
elucidated. Magnetothermal effects do not seem to be present in this material.

%

\section{Transport properties of nonintegrable systems}
\label{sec:3}

While originally conjectured to exhibit diffusive transport
properties \cite{castella95,zotos96,rosch00}, upon the experimental observation of 
large thermal conductivities in spin ladder, nonintegrable models have been discussed controversially in the 
literature
\cite{alvarez02b,heidarian07,hm02,hm03,narozhny98,alvarez02,rabson04,mukerjee06,mukerjee07,hm04c,orignac03,saito03,zotos04,sakai_yam05}. As of now, many studies point at 
a vanishing of both Drude weights \cite{heidarian07,hm02,hm03,mukerjee07,hm04c,shimshoni03,jung06} in massive phases 
of nonintegrable models, including spin ladders. The massless regime of the frustrated chain remains a controversial issue
\cite{heidarian07,hm03}. Here we illustrate some numerical results for the finite-size scaling of the Drude weights
\cite{hm03} taking the example
of the frustrated and dimerized chain and in particular, we also discuss the frequency dependence of the transport coefficients.

The Hamiltonian of the dimerized and frustrated spin-$1/2$ chain is:
\begin{equation}
H=\sum_{l=1}^{N} h_l=J\sum_{l=1}^{N}\,\lbrack \, \lambda_l \vec{S}_l \cdot\vec{S}_{l+1} + \alpha \vec{S}_l\cdot \vec{S}_{l+2}\rbrack\,;
\label{eq:f1}
\end{equation}
where $\alpha$ parameterizes the frustration and dimerization is introduced through 
 $\lambda_l=1(\lambda)$ for an even(odd) site index $l$ ($\lambda \leq 1$). 
The current operators derive from Eq.~(\ref{eq:f2}) \cite{hm02}.
The regular part of $\kappa[\sigma](\omega)$ appearing in Eq.~(\ref{eq:defineD}) can be written as:
\begin{equation}
\kappa[\sigma]_{\mathrm{reg}}(\omega) =
   \frac{\pi\beta^r }{Z\,N} \frac{1-e^{-\beta \omega}}{\omega}\sum_{m, n\atop E_m\not= E_n} e^{-E_n/T}
      |\langle m | j_{\mathrm{th[s]}}|n\rangle|^2 \,\delta(\omega-(E_m-E_n)) \,.
\label{eq:4}
\end{equation}

\subsection{Frustrated chain}
\label{sec:3a}

\begin{figure}
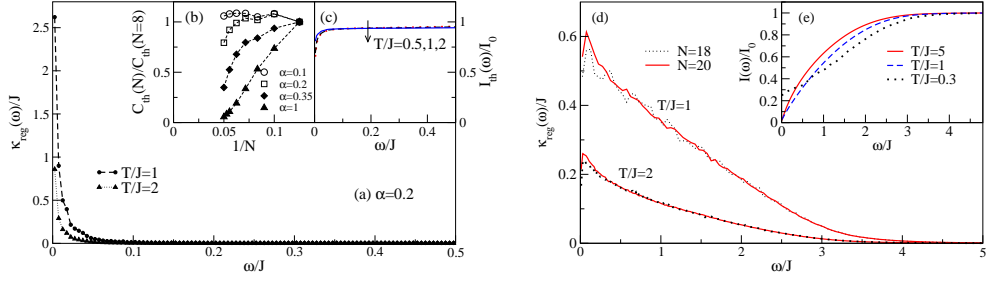

\centerline{\epsfig{figure=figure1.eps,height=0.25\textwidth}\hspace{0.4cm} \epsfig{figure=figure1a.eps,height=0.25\textwidth}}
\caption{
(a):  Regular part of the thermal conductivity $\kappa_{\mathrm{reg}}(\omega)$  for a frustrated chain with $\alpha=0.2$ [$N=18$
sites; $T/J=1,2$ (circles, triangles)]. (b): Finite-size scaling of the thermal Drude weight in the high-temperature limit:
$C_{\mathrm{th}}=\lim_{T\to\infty} \lbrack T^2 D_{\mathrm{th}}\rbrack$ for $\alpha=0.1,0.2,0.35,1$ (circles, squares, diamonds, triangles) for
$N=8,10,\dots,20$ sites (see also Ref.~\cite{hm03}). 
(c): Integrated spectral weight $I_{\mathrm{th}}$ for $\alpha=0.2$ and $T/J=0.5,1,2$.  
(d):  $\kappa_{\mathrm{reg}}(\omega)$ as a function of frequency $\omega$ for  $\alpha=1$ [$N=18,20$].
(e): Integrated spectral weight vs frequency at $T/J=0.3,1,5$ ((d),(e): reproduced from Ref.~\cite{hm04c}).
}\label{fig:1}
\end{figure}

As a result of preceding studies of the finite-size scaling of the thermal Drude
weight \cite{hm03}, we 
concluded that no indications for a finite Drude weight are evident from the system sizes accessible by ED.
This result is illustrated in Fig.~\ref{fig:1}(b), where we show the leading coefficient of an expansion of the thermal Drude weight 
$D_{\mathrm{th}}(T)$ in powers of $1/T$, i.e., $C_{\mathrm{th}}=\lim_{T\to\infty}\lbrack T^2 D_{\mathrm{th}}(T)\rbrack$, as a 
function of system size
(including new data for $N=20$ as compared to Ref.~\cite{hm03}). The decrease of $C_{\mathrm{th}}$  with $N$ is evident 
for all $\alpha$ as soon as the system size $N$ becomes large enough
 (see Ref.~\cite{hm03} for details).
The same picture arises for spin transport \cite{hm03}.\\\indent
Let us mention  the main features of $\mbox{Re}\,\kappa(\omega)$ as found for the case
of $\alpha=1$, i.e., in the massive regime, shown in Fig.~\ref{fig:1}(d) \cite{hm04c}: (i) $\kappa_{\mathrm{reg}}(\omega)$ is a broad,
 featureless function extending up to frequencies $\omega/J\lesssim 4$; (ii) at $T/J=1$ and $N=20$, the thermal Drude weight 
 only gives a small contribution
 to the  total weight of less than 3\%.

 We now proceed by a discussion of the frequency dependence of the 
thermal conductivity of frustrated chains in the massless regime, i.e.,
$\alpha \lesssim 0.241$ \cite{nomura93}.
Our numerical results for $\kappa_{\mathrm{reg}}(\omega)$ and $\alpha=0.2$ are shown in Fig.~\ref{fig:1}(a) for $N=18$ sites and
$T/J=1,2$.  $\kappa_{\mathrm{reg}}(\omega)$ consists of a narrow peak centered around $\omega=0$, extending up to $\omega/J\lesssim 0.05$. 
This is 
reflected in the integrated spectral weight $I_{\mathrm{th}}(\omega)=\int_0^{\omega} d\omega'\,
\mbox{Re}\kappa(\omega')$,  depicted in Fig.~\ref{fig:1}(c).
As this quantity  also includes the contribution from the Drude weight, the figure reveals that the thermal Drude weight {\it on the system sizes considered here}
 amounts to  more than
 50\% of the total spectral weight. This observation is in stark contrast to the behavior of 
 $\kappa(\omega)$ in the massive regime on chains of a comparable length as summarized above.  While on the one hand, the analysis of the finite-frequency properties of $\kappa$ for $\alpha > 0.241$ 
 supports the conclusion of a vanishing Drude weight, the question arises on the other hand whether the conclusion of $D_{\mathrm{th}}\to 0$ 
 needs to be  reconsidered in the massless regime. 

 From the numerical data, it is  difficult to establish a definitive conclusion about the thermodynamic limit from 
 small system sizes as far as current-current correlation functions are concerned. Still, no substantial differences 
 are found between the massless and the massive regime concerning the finite-size scaling of the thermal
 Drude weight \cite{hm03}, the 
 common feature being a monotonic decrease of $D_{\mathrm{th}}$ with system size $N$ for $N$ large enough. Moreover, bosonization studies on general grounds predict
 a vanishing Drude weight for thermal transport, irrespective of the presence of a
 gap \cite{hm02,hm03,rosch00,shimshoni03}.
 In a recent work \cite{jung06}, Jung {\it et al.}~have shown that, to first order in $\alpha$, the commutator 
 $\lbrack H,j_{\mathrm{th}} \rbrack$ between the Hamiltonian and the energy current operator  vanishes, preserving
 the exact conservation of the energy current operator of the nearest-neighbor $XXZ$ chain. While this feature explains the peculiar behavior
 of $\kappa(\omega)$ for small system sizes as observed here, a vanishing of the Drude weight can still
 be expected in the thermodynamic
 limit where {\it all} terms of $\lbrack  H, j_{\mathrm{th}} \rbrack$  in powers of $\alpha$ become relevant and 
 cause a finite dc conductivity. The results of a quantum Monte-Carlo (QMC) study, however, seem to indicate that in massless phases 
 of nonintegrable models, finite Drude weights may exist \cite{heidarian07}. Note though that the interpretation of Monte-Carlo
 data at finite frequencies is quite involved as an analytic continuation from Matsubara to real frequencies needs to be
 performed \cite{kirchner99,heidarian07,alvarez02}. \\\indent
 In summary, the peculiar feature of $\kappa(\omega)$ of frustrated chains in the massless regime, i.e. $\alpha<0.241$, as 
 found for the system sizes accessible numerically, is that
 most spectral weight is found in the Drude weight, 
 while the regular part consists of a narrow peak around $\omega=0$  only.

 \begin{figure}[t]
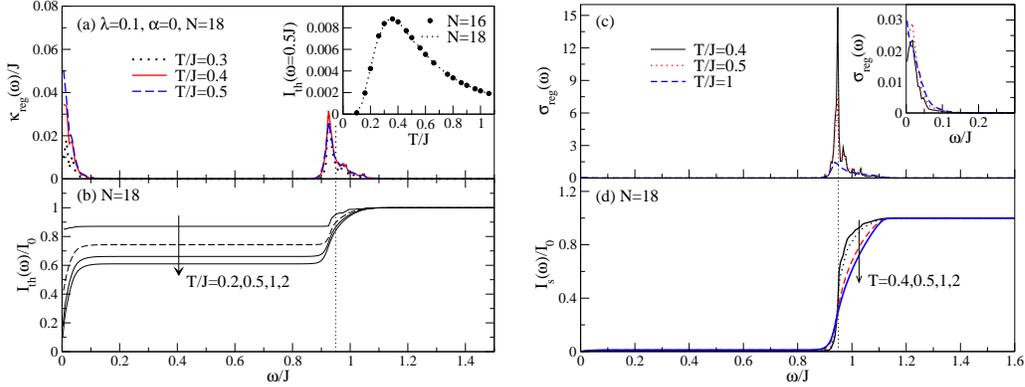

\centerline{\epsfig{figure=figure2.eps,width=0.45\textwidth}\hspace{0.4cm} \epsfig{figure=figure3.eps,width=0.45\textwidth}}
\caption{Dimerized chain with $\lambda=0.1$. (a): Regular part of the thermal conductivity as a function of frequency $\omega$  ($N=18$
sites; $T/J=0.3,0.4,0.5$; solid, dotted, dashed line). (b): Integrated weight $I_{\mathrm{th}}$ vs $\omega$ for
$T/J=0.2,0.5,1,2$.
Inset of (a): Integrated weight $I_{\mathrm{th}}(\omega)$ for $\omega/J=0.5$ as a function of temperature.
(c): Regular part of the spin conductivity vs $\omega$   ($N=18$
sites; $T/J=0.4,0.5,1$; solid, dotted, dashed line). (d): Integrated weight $I_{\mathrm{s}}$ vs $\omega$ for
$T/J=0.4,0.5,1,2$.
Inset of (c): Enlarged view of the low-frequency region of panel (c).
Vertical, dotted lines mark the position of the spin gap.}\label{fig:2}
\end{figure}

\subsection{The dimerized chain}

We next address finite-frequency transport properties
of the dimerized chain ($\alpha=0$).
In the following, we choose $\lambda=0.1$, i.e., we focus  on the limit of strong dimerization $\lambda \ll 1$.  
To first order in $\lambda$, the dispersion relation of the elementary triplet excitation is 
described by $\epsilon_k/J = 1+(\lambda/2) \cos(k)$ \cite{harris73},
where $k$ denotes the momentum. The spin gap $G$ is quite large and roughly given by
$G/J=0.95$, while 
triplet-triplet interactions are suppressed  by decreasing $\lambda$.  One may therefore on the one hand expect both the spin and heat conductivity to be small
due to the large spin gap, but on the other hand, the transport properties should be well approximated by considering a weakly interacting
gas of hardcore bosons  \cite{sachdev90}, which may, as a future project, allow for a comparison between 
numerical and analytical results.\\
\indent
Our numerical results for the  conductivities $\kappa(\omega)$ and $\sigma(\omega)$ are presented in Figs.~\ref{fig:2}(a) and \ref{fig:2}(c), respectively.
The computations were performed for $N=18$ sites, $\lambda=0.1$, and several finite temperatures as listed in the figure's caption.
The distinctive features of both conductivities visible in Figs.~\ref{fig:2}(a) and \ref{fig:2}(c) are:
(i) Significant spectral weight is only found around $\omega=0$ and in a high-frequency peak located
around $\omega/J\gtrsim  0.95$, which corresponds to the spin gap.
(ii)  While the low 
frequency peak (including the Drude weight) contains a large fraction of the total weight in the case of the thermal conductivity,
the spectral weight of the spin conductivity is mainly concentrated in the high-frequency peak. The latter is 
illustrated in Figs.~\ref{fig:2}(b) and \ref{fig:2}(d), showing the integrated spectral weight $I_{\mathrm{th[s]}}(\omega)/I_{0}$, where $I_0$ is the
full spectral weight of $\kappa[\sigma](\omega)$.
The low-frequency peak is present in $\sigma(\omega)$ as well.
The inset 
and the  main panel of Fig.~\ref{fig:2}(c) show that the low-frequency
peak extends up to $\omega/J\sim 0.1$, which 
corresponds to the width of the one-triplet band.\\
\indent
 Furthermore, by integrating the low-frequency peak in $\kappa_{\mathrm{reg}}(\omega)$ over $\omega$ up to 
$\omega/J\approx 0.5$ yielding $I_{\mathrm{reg}}(\omega/J=0.5)$, we find that this quantity is independent of system size within numerical 
precision. 
$I_{\mathrm{th}}(\omega/J=0.5)$  is plotted in the inset of Fig.~\ref{fig:2}(a) for $N = 16$
and $N = 18$ sites. Hence, a significant redistribution of spectral weight as the system size increases is  not expected. 
One further observes a maximum in $I_{\mathrm{th}}(\omega/J=0.5)$ at roughly $T/J\sim 0.35$ and a $1/T^2$-dependence
at high temperatures.

In summary, both models exhibit an intriguing behavior of the frequency dependence of both the spin and thermal conductivity that
deserves further investigations.

\section{Extrinsic scattering}
\label{sec:ex}
As mentioned in the introduction, and as is evident form the phenomenological analysis of
experimental data for spin ladder \cite{solo00,hess01,kudo01,hess04,hess06} as well as spin chain materials 
\cite{solo01,hess07,solo03,ribeiro04},
it is important to include external scattering processes to arrive at a realistic theory of thermal transport in quasi 1D
magnetic materials. For instance, doping with nonmagnetic impurities in (Sr,Ca,La)$_{14}$Cu$_{24}$O$_{41}$ \cite{hess06}
-- substitution of Zn for Cu --
has been found to result in a suppression of the thermal conductivity
linear in the Zn content. Mobile charge carriers effectively suppress the magnon thermal transport in spin ladder systems
\cite{hess04}. As heat transport via magnetic systems in a material requires the heat to be transferred from the lattice
to the spin system, inevitably, spin-phonon scattering needs to be modeled  by theory. 

First studies have addressed the thermal conductivity of spin-phonon coupled spin chains
\cite{shimshoni03,rozhkov05a,chernychev05,louis06} as well as spin ladders \cite{boulat06}. Some of these works
\cite{shimshoni03,boulat06} start from effective field theories and describe transport within the Memory-matrix
formalism \cite{forster,jung07} by first identifying the slowest decaying modes, following the spirit of Ref.~\cite{rosch00}.  These then determine
the long-time behavior of current-current correlation functions. For the case of spin chains, an exponentially large 
thermal conductivity $\kappa_{\mathrm{total}}\propto \mbox{exp}(a\,\Theta/2T)$ is predicted  \cite{shimshoni03}, where $\Theta$ is
the Debye temperature. A peculiar result of the Boltzmann theory
of Refs.~\cite{rozhkov05a,chernychev05} is the constant 
spin thermal conductivity at high temperatures.
For spin ladders, Ref.~\cite{boulat06} highlights the relevance of spin-phonon drag terms contributing to the total
thermal conductivity, with a rich interplay of energy scales influencing the low-temperature behavior. A direct comparison of these
results with
experiments, however, needs to be done in future, in particular, as disorder may be of relevance in the structurally disordered
spin ladder compounds (Sr,Ca,La)$_{14}$Cu$_{24}$O$_{41}$ \cite{solo00,hess01,boulat06}. 
Finally, note that spin phonon coupling has also been studied in the context of spin transport in the spin-1/2 chain by means of 
QMC \cite{louis05}.

Note
that via the Jordan-Wigner transformation Heisenberg type of models can be mapped onto
spinless fermions \cite{mahan}, the transport properties of which have extensively been
studied in the context of localization
\cite{kramer93}. We
just mention an incomplete list of recent, closely related works addressing Heisenberg chains 
\cite{bouzerar94,motrunich00,motrunich01,byrnes02,carvalhodias05,hm06}, spin ladders \cite{orignac03}, or effective low-energy models 
\cite{rozhkov05a,chernychev05,kane96}.  Interestingly, some works seem to
indicate that the dc spin conductivity may be finite for interacting systems in the case of off-diagonal disorder 
\cite{motrunich00,motrunich01}. Also, even if the dc spin conductivity vanishes, the same is not necessarily true for 
thermal transport as energy can still be transfered over a weak link \cite{kane96}.
 
Finally, only results from a mean-field
theory are available for  the thermal conductivity of doped spin ladders
in the literature \cite{qin04}. Transport properties of 1D $t$-$J$ and Hubbard models 
have widely been investigated (see, e.g., Ref.~\cite{zotos-review} for
an overview), and it is beyond the scope of this work to discuss the charge
and spin transport of these systems. Their thermal transport properties
have, however, not been
studied sufficiently \cite{peterson07}. Note that the Hubbard model, being integrable, is expected to exhibit ballistic 
thermal transport, which also holds
for the supersymmetric point of the  $t$-$J$ model \cite{zotos97}.

\section{Summary}
\label{sec:conc}

We may conclude that the intrinsic thermal transport properties of the spin-1/2 $XXZ$ chain in zero and finite longitudinal
fields are well understood. The spin transport of this model still poses some challenges to theorists, such as 
an analytical calculation of the spin Drude weight of the  spin-1/2 Heisenberg chain. As for nonintegrable systems and
within linear response theory, it seems that generically, ballistic transport in the sense of finite Drude weights is not
realized. Rather, the relevant information is encoded in the frequency dependence of the conductivities. The challenge to 
computational scientists is to devise algorithms that can simulate low temperature regimes. Analytical approaches face the 
problem that effective field-theories of nonintegrable models are typically integrable, with diverging transport coefficients. 
Hence, the definition of a low-energy  theory that describes transport accurately is a nontrivial task.
 
Promising results with respect to the interpretation of experiments have been obtained from first studies
incorporating phonons or disorder, but a consistent picture has not emerged yet.

Highly interesting and potentially new physics is expected  from both experiments and novel theoretical methods 
such as the time-dependent density matrix renormalization group method \cite{white04}
that investigate transport and relaxation of strongly-correlated electron systems away from equilibrium.

{\bf Acknowledgments} This work has been possible only through collaborations and discussions with B.\ B\"uchner, D.C.\ Cabra,
and C.\ Hess, which we gratefully acknowledge.\ We would also like to thank 
N.\ Andrei, J.\ Gemmer, C.\ Gros, P.\ Jung, A.\ Kl\"umper, T.\ Lorenz, K.\ Louis,
M.\ Michel, A.\ Rosch, A.\ Sologubenko, and X.\ Zotos for fruitful discussions.



\begin{thebibliography}{100}

\bibitem{dagotto99}
E.~Dagotto,
\newblock Rep.\ Prog.\ Phys. {\bf 62} (1999) 1525.

\bibitem{johnston00}
D.~C. Johnston, M.~Troyer, S.~Miyahara, D.~Lidsky, K.~Ueda, M.~Azuma, Z.~Hiroi,
  M.~Takano, M.~Isobe, Y.~Ueda, M.~A. Korotin, V.~I. Anisimov, A.~V. Mahajan,
  and L.~L. Miller,
\newblock cond-mat/0001147  (unpublished);
D.~C. Johnston, R.~K. Kremer, M.~Troyer, X.~Wang, A.~Kl\"umper, S.~L. Bud'ko,
  A.~F. Panchula, and P.~C. Canfield,
\newblock Phys.\ Rev.\ B {\bf 61} (2000) 9558.

\bibitem{schulz96}
H.~Schulz,
\newblock {\em in: Correlated Fermions and Transport in Mesoscopic Systems},
  edited by T. Martin, G. Montambaux, and J. Tran Thanh Van,
\newblock Editions Fronti\`eres, Gif-sur-Yvette, 1996.

\bibitem{bethe31}
H.~Bethe,
\newblock Z. Phys. {\bf 71} (1931) 205;
A.~Kl\"umper,
\newblock Lect. Notes Phys. {\bf 645} (2004) 349.

\bibitem{white92b}
S.~R. White,
\newblock Phys.\ Rev.\ Lett. {\bf 69} (1992) 2863;
U.~Schollw\"ock,
\newblock Rev. Mod. Phys. {\bf 77} (2005) 259.


\bibitem{qm-review}
U.~Schollw\"ock, J.~Richter, D.~Farnell, and R.~Bishop, editors,
\newblock {\em 
 Quantum Magnetism}, volume 645 of {\em Lecture Notes in  Physics},
\newblock Springer, Berlin Heidelberg New York, 2004.

\bibitem{takigawa96}
M.~Takigawa, N.~Motoyama, H.~Eisaki, and S.~Uchida,
\newblock Phys.\ Rev.\ Lett. {\bf 76} (1996) 4612;
K.~Thurber, A.~Hunt, T.~Imai, and F.~Chou,
\newblock Phys. Rev. Lett. {\bf 87} (2001) 247202.

\bibitem{mahan}
G.~D. Mahan,
\newblock {\em Many-{P}article {P}hysics},
\newblock Plenum Press, New York London, 1990.



\bibitem{solo00}
A.~V. Sologubenko, K.~Gianno, H.~R. Ott, U.~Ammerahl, and A.~Revcolevschi,
\newblock Phys.\ Rev.\ Lett. {\bf 84} (2000) 2714.

\bibitem{hess01}
C.~Hess, C.~Baumann, U.~Ammerahl, B.~B\"uchner, F.~Heidrich-Meisner, W.~Brenig,
  and A.~Revcolevschi,
\newblock Phys.\ Rev.\ B {\bf 64} (2001) 184305.

\bibitem{kudo01}
K.~Kudo, S.~Ishikawa, T.~Noji, T.~Adachi, Y.~Koike, K.~Maki, S.~Tsuji, and
  K.~Kumagai,
\newblock J. Phys. Soc. Jpn. {\bf 70} (2001) 437.

\bibitem{alvarez02b}
J.~V. Alvarez and C.~Gros,
\newblock Phys.\ Rev.\ Lett. {\bf 89 }(2002) 156603.

\bibitem{solo01}
A.~V. Sologubenko, K.~Gianno, H.~R. Ott, A.~Vietkine, and A.~Revcolevschi,
\newblock Phys.\ Rev.\ B {\bf 64} (2001) 054412; 
A.~V. Sologubenko, E.~Felder, K.~Gianno, H.~R. Ott, A.~Vietkine, and
  A.~Revcolevschi,
\newblock Phys.\ Rev.\ B {\bf 62} (2000) R6108.

\bibitem{hess07}
C.~Hess, H.~ElHaes, A.~Waske, B.~B\"uchner, C.~Sekar, G.~Krabbes,
  F.~Heidrich-Meisner, and W.~Brenig,
\newblock Phys.\ Rev.\ Lett. {\bf 98} (2007) 027201.

\bibitem{sales02}
B.~C. Sales, M.~D. Lumsden, S.~E. Nagler, D.~Mandrus, and R.~Jin,
\newblock Phys. Rev. Lett. {\bf 88} (2002) 095901; 
B.~C. Sales, R.~Jin, and D.~Mandrus,
\newblock cond-mat/0401154.

\bibitem{sun03}
X.~F. Sun, S.~Komiya, and Y.~Ando,
\newblock Phys.\ Rev.\ B {\bf 67} (2003) 184512;
M.~Hofmann, T.~Lorenz, K.~Berggold, M.~Gr\"uninger, A.~Freimuth, G.~S. Uhrig,
  and E.~Br\"uck,
\newblock Phys.\ Rev.\ B {\bf 67} (2003) 184502;
R.~Jin, Y.~Onose, Y.~Tokura, D.~Mandrus, P.~Dai, and B.~C. Sales,
\newblock Phys. Rev. Lett. {\bf 91} (2003) 146601;
C. Hess, B. B\"uchner,  U. Ammerahl, L. Colonescu,   F. Heidrich-Meisner,  W. Brenig, and A. Revcolevschi,
\newblock Phys. Rev. Lett. {\bf 90} (2003) 197002.

\bibitem{hofmann01}
M.~Hofmann, T.~Lorenz, G.~S. Uhrig, H.~Kierspel, O.~Zabara, A.~Freimuth,
  H.~Kageyama, and Y.~Ueda,
\newblock Phys.\ Rev.\ Lett. {\bf 87} (2001) 047202.

\bibitem{ando98}
Y.~Ando, J.~Takeya, D.~L. Sisson, S.~G. Doettinger, I.~Tanaka, R.~S. Feigelson,
  and A.~Kapitulnik,
\newblock Phys.\ Rev.\ B {\bf 58} (1998) R2913;
J.~Takeya, I.~Tsukada, Y.~Ando, T.~Masuda, and K.~Uchinokura,
\newblock Phys.\ Rev.\ B {\bf 62} (2000) R9260;
J.~Takeya, I.~Tsukada, Y.~Ando, T.~Masuda, K.~Uchinokura, I.~Tanaka, R.~S.
  Feigelson, and A.~Kapitulnik,
\newblock Phys.\ Rev.\ B {\bf 63} (2001) 214407.

\bibitem{sologubenko07}
A.~V. Sologubenko, K.~Berggold, T.~Lorenz, A.~Rosch, E.~Shimshoni, M.~D.
  Phillips, and M.~M. Turnbull,
\newblock Phys. Rev. Lett. {\bf 98} (2007) 107201.





\bibitem{sologubenko07a}
A.~V. Sologubenko, T.~Lorenz, H.~R. Ott, and A.~Freimuth,
\newblock J. Low Temp. Phys. {\bf 147} (2007) 387.


\bibitem{luttinger64}
J.~M. Luttinger,
\newblock Phys.\ Rev. {\bf 135} (1964) A1505;
B.~Shastry,
\newblock Phys.\ Rev.\ B {\bf 73} (2006) 085117.

\bibitem{saito96}
K.~Saito, S.~Takesue, and S.~Miyashita,
\newblock Phys.\ Rev.\ E {\bf 54} (1996) 2404;
K.~Saito,
\newblock Europhys.\ Lett. {\bf 61} (2002) 34;
K.~Saito and S.~Miyashita,
\newblock J. Phys. Soc. Jpn. {\bf 71} (2002) 2485.

\bibitem{mejia-monasterio06}
C.~Mejía-Monasterio, T.~Prosen, and G.~Casati,
\newblock Europhys. Lett. {\bf 72} (2006) 520.

\bibitem{michel03}
M.~Michel, M.~Hartmann, J.~Gemmer, and G.~Mahler,
\newblock Eur. Phys. J B {\bf 34} (2003) 325.

\bibitem{michel04}
M.~Michel, J.~Gemmer, and G.~Mahler,
\newblock Eur. Phys. J B {\bf 42} (2004) 555.

\bibitem{michel05a}
M.~Michel, G.~Mahler, and J.~Gemmer,
\newblock Phys. Rev. Lett. {\bf 95} (2005) 180602.

\bibitem{gemmer06}
J.~Gemmer, R. Steinigeweg, and M. Michel,
\newblock Phys. Rev. B {\bf 73} (2006) 104302.

\bibitem{sachdev97}
S.~Sachdev and K.~Damle,
\newblock Phys.\ Rev.\ Lett. {\bf 78} (1997) 943;
K.~Damle and S.~Sachdev,
\newblock Phys.\ Rev.\ B {\bf 57} (1998) 8307;
S.~Fujimoto,
\newblock J. Phys. Soc. Jpn. {\bf 68} (1999) 2810;
J.~Karadamoglou and X.~Zotos,
\newblock Phys.\ Rev.\ Lett. {\bf 93} (2004) 177203.

\bibitem{hida92}
K.~Hida,
\newblock Phys.\ Rev.\ B {\bf 45} (1992) 2207;
S.~R. White,
\newblock Phys. Rev. B {\bf 53} (1996) 52;
A.~Kolezhuk and H.~Mikeska,
\newblock Phys. Rev. B {\bf 56} (1997) R11380.

\bibitem{lepri03}
S.~Lepri, R.~Livi, and A.~Politi,
\newblock Phys. Rep. {\bf 377} (2003) 1.

\bibitem{kohn64}
W.~Kohn,
\newblock Phys. Rev {\bf 133} (1964) A171.

\bibitem{scalapino93}
D.~J. Scalapino, S.~R. White, and S.~Zhang,
\newblock Phys.\ Rev.\ B {\bf 47} (1993) 7995.

\bibitem{zotos-review}
X.~Zotos and P.~Prelov{\v{s}}ek,
\newblock {\em in: Strong Interactions in Low Dimensions}, chapter~11,
\newblock Physics and Chemistry of Materials with Low-Dimensional Structures,
  Kluwer Academic Publishers, 2004.

\bibitem{niemeyer71}
T.~Niemeyer and H.~van Vianen,
\newblock Phys. Lett. {\bf 34A} (1971) 401.

\bibitem{zotos97}
X.~Zotos, F.~Naef, and P.~Prelov{\v{s}}ek,
\newblock Phys.\ Rev.\ B {\bf 55} (1997) 11029.

\bibitem{zotos05}
X.~Zotos,
\newblock J. Phys. Soc. Jpn. Suppl. {\bf 74} (2005) 173.

\bibitem{castella95}
H.~Castella, X.~Zotos, and P.~Prelov{\v{s}}ek,
\newblock Phys.\ Rev.\ Lett. {\bf 74} (1995) 972.

\bibitem{zotos96}
X.~Zotos and P.~Prelov{\v{s}}ek,
\newblock Phys.\ Rev.\ B {\bf 53} (1996) 983.



\bibitem{kirchner99}
S.~Kirchner, H.~G. Evertz, and W.~Hanke,
\newblock Phys.\ Rev.\ B {\bf 59} (1999) 1825.

\bibitem{heidarian07}
D.~Heidarian and S.~Sorella,
\newblock Phys. Rev. B {\bf 75} (2007) 241104(R).

\bibitem{kluemper02}
A.~Kl\"umper and K.~Sakai,
\newblock J. Phys. A {\bf 35} (2002) 2173.

\bibitem{sakai03}
K.~Sakai and A.~Kl\"umper,
\newblock J. Phys. A {\bf 36}  (2003) 11617.

\bibitem{hm02}
F.~Heidrich-Meisner, A.~Honecker, D.~C. Cabra, and W.~Brenig,
\newblock Phys.\ Rev.\ B {\bf 66} (2002) 140406(R).

\bibitem{hm03}
F.~Heidrich-Meisner, A.~Honecker, D.~C. Cabra, and W.~Brenig,
\newblock Phys. Rev. B {\bf 68} (2003) 134436;
\newblock J. Mag. Mag. Mat. {\bf 272-276}  (2004) 890;
\newblock Phys. Rev. Lett. {\bf 92} (2004) 069703.


\bibitem{lou04}
P.~Lou, W.-C. Wu, and M.-C. Chang,
\newblock Phys.\ Rev.\ B {\bf 70} (2004) 064405.


\bibitem{narozhny98}
B.~N. Narozhny, A.~J. Millis, and N.~Andrei,
\newblock Phys.\ Rev.\ B {\bf 58} (1998) R2921.

\bibitem{fabricius98}
K.~Fabricius and B.~M. McCoy,
\newblock Phys.\ Rev.\ B {\bf 57} (1998) 8340.

\bibitem{naef98}
F.~Naef and X.~Zotos,
\newblock J. Phys. C {\bf 10} (1998) L183.

\bibitem{zotos99}
X.~Zotos,
\newblock Phys.\ Rev.\ Lett. {\bf 82} (1999) 1764.

\bibitem{peres99}
N.~M.~R. Peres, P.~D. Sacramento, D.~K. Campbell, and J.~M.~P. Carmelo,
\newblock Phys.\ Rev.\ B {\bf 59} (1999) 7382;
S.-J. Gu, V.~M. Pereira, and N.~M.~R. Peres,
\newblock Phys.\ Rev.\ B {\bf 66} (2002) 235108.

\bibitem{alvarez02}
J.~V. Alvarez and C.~Gros,
\newblock Phys.\ Rev.\ Lett. {\bf 88} (2002) 077203;
\newblock Phys.\ Rev.\ B {\bf 66} (2002) 094403.

\bibitem{fujimoto03}
S.~Fujimoto and N.~Kawakami,
\newblock Phys.\ Rev.\ Lett. {\bf 90} (2003) 197202.


\bibitem{long03}
M.~W. Long, P.~Prelov{\v{s}}ek, S.~ElShawish, J.~Karadamoglou, and X.~Zotos,
\newblock Phys. Rev. B {\bf 68} (2003) 235106.
\bibitem{rabson04}
D.~A. Rabson, B.~N. Narozhny, and A.~J. Millis,
\newblock Phys. Rev. B {\bf 69} (2004) 054403.

\bibitem{laflorencie04}
N.~Laflorencie and H.~Rieger,
\newblock Eur. Phys. J. B {\bf 40} (2004) 201.

\bibitem{prelovsek04}
P.~Prelov{\v{s}}ek, S.~ElShawish, X.~Zotos, and M.~W. Long,
\newblock Phys.\ Rev.\ B {\bf 70} (2004) 205129.

\bibitem{benz05}
J.~Benz, T.~Fukui, A.~Kl\"umper, and C.~Scheeren,
\newblock J. Phys. Soc. Jpn. Suppl. {\bf 74} (2005) 181.

\bibitem{gobert05}
D.~Gobert, C.~Kollath, U.~Schollw\"ock, and G.~Sch\"utz,
\newblock Phys.\ Rev.\ E {\bf 71} (2005) 036102.

\bibitem{sirker06}
J.~Sirker,
\newblock Phys. Rev. B {\bf 73} (2006) 224424.

\bibitem{mukerjee06}
S.~Mukerjee, V.~Oganesyan, and D.~Huse,
\newblock Phys. Rev. B {\bf 73} (2006) 035113.

\bibitem{mukerjee07}
S.~Mukerjee and B.~S. Shastry,
\newblock arXiv:0705.3791.

\bibitem{shastry90}
B.~Shastry and B.~Sutherland,
\newblock Phys.\ Rev.\ Lett. {\bf 65} (1990) 243.

\bibitem{lan07}
Z.~Qiu-Lan and G.~Shi-Jian,
\newblock Chinese Physics Letters {\bf 24} (2007) 1354.

\bibitem{hm-diss}
F.~Heidrich-Meisner,
\newblock {\em Transport properties of low-dimensional quantum spin systems},
\newblock Ph.D. thesis, Technische Universit\"at Braunschweig,
  http://opus.tu-bs.de/opus/volltexte/2005/712/, 2005.

\bibitem{louis03}
K.~Louis and C.~Gros,
\newblock Phys.\ Rev.\ B {\bf 67} (2003) 224410.

\bibitem{hm05}
F.~Heidrich-Meisner, A.~Honecker, and W.~Brenig,
\newblock Phys. Rev. B {\bf 71} (2005) 184415.

\bibitem{sakai05}
K.~Sakai and A.~Kl\"umper,
\newblock J. Phys. Soc. Jpn. Suppl. {\bf 74} (2005) 74.

\bibitem{furukawa05}
S.~Furukawa, D.~Ikeda, and K.~Sakai,
\newblock J. Phys. Soc. Jpn. {\bf 74} (2005) 3241.

\bibitem{cabra98}
D.~C. Cabra, A.~Honecker, and P.~Pujol,
\newblock Phys.\ Rev.\ B {\bf 58} (1998) 6241.

\bibitem{hammar99}
P.~R. Hammar, M.~B. Stone, D.~H. Reich, C.~Broholm, P.~J. Gibson, M.~M.
  Turnbull, C.~P. Landee, and M.~Oshikawa,
\newblock Phys.\ Rev.\ B {\bf 59} (1999) 1008.

\bibitem{rosch00}
A.~Rosch and N.~Andrei,
\newblock Phys.\ Rev.\ Lett. {\bf 85} (2000) 1092.


\bibitem{hm04c}
F.~Heidrich-Meisner, A.~Honecker, D.~C. Cabra, and W.~Brenig,
\newblock Physica B {\bf 359-361} (2005) 1394.

\bibitem{orignac03}
E.~Orignac, R.~Chitra, and R.~Citro,
\newblock Phys.\ Rev.\ B {\bf 67} (2003) 134426.

\bibitem{saito03}
K.~Saito,
\newblock Phys. Rev. B {\bf 67} (2003) 064410.

\bibitem{zotos04}
X.~Zotos,
\newblock Phys. Rev. Lett. {\bf 92}  (2004) 067202.

\bibitem{sakai_yam05}
T.~Sakai and S.~Yamamoto,
\newblock J. Phys. Soc. Jpn. Suppl. {\bf 74} (2005) 191.

\bibitem{shimshoni03}
E.~Shimshoni, N.~Andrei, and A.~Rosch,
\newblock Phys. Rev. B {\bf 68} (2003) 104401.

\bibitem{jung06}
P.~Jung, R.~W. Helmes, and A.~Rosch,
\newblock Phys.\ Rev.\ Lett. {\bf 96} (2005) 067202.


\bibitem{nomura93}
K.~Nomura and K.~Okamoto,
\newblock J. Phys. Soc. Jpn. {\bf 62} (1993) 1123.

\bibitem{harris73}
A.~B. Harris,
\newblock Phys.\ Rev.\ B {\bf 7} (1973) 3166;
G.~S. Uhrig,
\newblock Phys. Rev. Lett. {\bf 79} (1997) 163.

\bibitem{sachdev90}
S. Sachdev and R.N. Bhatt,
\newblock Phys.\ Rev.\ B {\bf 41} (1990) 9323.

\bibitem{hess06}
C.~Hess, P.~Ribeiro, B.~B\"uchner, H.~ElHaes, G.~Roth, U.~Ammerahl, and
  A.~Revcolevschi,
\newblock Phys. Rev. B {\bf 73} (2006) 104407.

\bibitem{hess04}
C.~Hess, H.~ElHaes, B.~B\"uchner, U.~Ammerahl, M.~H\"ucker, and
  A.~Revcolevschi,
\newblock Phys. Rev. Lett. {\bf 93} (2004) 027005;
C.~Hess, C.~Baumann, and B.~B\"uchner,
\newblock J. Mag. Mag. Mat. {\bf 290-291} (2005) 322.




\bibitem{solo03}
A.~V. Sologubenko, H.~R. Ott, G.~Dhalenne, and A.~Revcolevschi,
\newblock Europhys. Lett. {\bf 62} (2003) 540.


\bibitem{ribeiro04}
P.~Ribeiro, C.~Hess, P.~Reutler, G.~Roth, and B.~B\"uchner,
\newblock J. Mag. Mag. Mat. {\bf 290-291} (2005) 334.


\bibitem{rozhkov05a}
A.~V. Rozhkov and A.~L. Chernyshev,
\newblock Phys.\ Rev.\ Lett. {\bf 94} (2005) 087201.

\bibitem{chernychev05}
A.~L. Chernyshev and A.~V. Rozhkov,
\newblock Phys. Rev. B {\bf 72} (2005) 104423.

\bibitem{louis06}
K.~Louis and P.~Prelov{\v{s}}ek and X.~Zotos,
\newblock Phys.\ Rev.\ B {\bf 74} (2006) 235118.

\bibitem{boulat06}
E.~Boulat, P.~Mehta, N.~Andrei, E.~Shimshoni, and A.~Rosch,
\newblock Phys. Rev. B {\bf 76} (2007) 214411.

\bibitem{forster}
D.~Forster,
\newblock {\em Hydrodynamic {F}luctuations, {B}roken {S}ymmetry and
  {C}orrelation {F}unctions},
\newblock Addison-Wesley Publishing, Reading, Massachusetts, 1990.

\bibitem{jung07}
P.~Jung and A.~Rosch,
\newblock Phys. Rev. B {\bf 75} (2007) 245104.

\bibitem{louis05}
K.~Louis and X.~Zotos,
\newblock Phys.\ Rev.\ B {\bf 72} (2005) 214415.


\bibitem{kramer93}
B.~Kramer and A.~MacKinnon,
\newblock Rep. Prog. Phys. {\bf 56} (1993) 1469.

\bibitem{bouzerar94}
G.~Bouzerar, D.~Poilblanc, and G.~Montambaux,
\newblock Phys. Rev. B {\bf 49} (1994) 8258.

\bibitem{motrunich00}
K.~Damle, O.~Motrunich, and D.~A. Huse,
\newblock Phys. Rev. B {\bf 84} (2000) 003434.

\bibitem{motrunich01}
O.~Motrunich, K.~Damle, and D.~A. Huse,
\newblock Phys. Rev. B {\bf 63} (2001) 134424.

\bibitem{byrnes02}
T.~M.~R. Byrnes, R.~J. Bursill, H.-P. Eckle, C.~J. Hamer, and A.~W. Sandvik,
\newblock Phys. Rev. B {\bf 66} (2002) 195313.

\bibitem{carvalhodias05}
F.~C. Dias, I.~R. Pimentel, and M.~Henkel,
\newblock Phys.\ Rev.\ B {\bf 73}, (2005) 075109.

\bibitem{hm06}
F.~Heidrich-Meisner,
\newblock Physica B {\bf 378} (2006) 299.

\bibitem{kane96}
C.~I. Kane and M.~P.~A. Fisher,
\newblock Phys.\ Rev.\ Lett. {\bf 76} (1996) 3192.

\bibitem{qin04}
J.~Qin, S.~Feng, F.~Yuan, and W.~Y. Chen,
\newblock Phys. Lett. A {\bf 335} (2005) 477.

\bibitem{peterson07}
M.R. Peterson, S. Mukerjee, B.S. Shastry, and J.O. Haerter,
\newblock Phys. Rev. B {\bf 76} (2007) 125110.


\bibitem{white04}
S.~R. White and A.~E. Feiguin,
\newblock Phys.\ Rev.\ Lett. {\bf 93} (2004) 076401;
A.~Daley, C.~Kollath, U.~Schollw\"ock, and G.~Vidal,
\newblock J. Stat. Mech.: Theory Exp. (2004) {\bf P04005};
K.~Al-Hassanieh, A.~Feiguin, J.~Riera, C.~B\"usser, and E.~Dagotto,
\newblock Phys. Rev. B {\bf 73} (2006) 195304;
G.~Schneider and P.~Schmitteckert,
\newblock cond-mat/0601389.

\end{thebibliography}
\end{document}